\documentclass[a4paper]{article}

\usepackage{a4wide}
\usepackage{amsmath,amssymb}
\usepackage[english]{babel}
\usepackage{cite}
\usepackage{url}
\usepackage[utf8x]{inputenc}
\usepackage{amsmath}
\usepackage{graphicx}
\usepackage{subfig}
\usepackage{booktabs}
\usepackage{multirow}
\usepackage[titletoc]{appendix}
\graphicspath{{images/}}
\usepackage{parskip}
\usepackage{fancyhdr}
\usepackage{hyperref}
\usepackage{kpfonts}
\usepackage{vmargin}
\usepackage{media9}
\usepackage{multimedia}
\usepackage{lmodern}
\begin{document}

\begin{center}
{\begin{Large}
\textbf{%
Data-driven modelling of scalable spinodoid structures for energy absorption}
\end{Large}}

\medskip

\underline{Hirak Kansara}$^{1}$, Gary Koh$^{1}$, Merrin Varghese$^{1}$, John Z. X. Luk$^{1}$, Emilio F. Gomez$^{1}$, Siddhant Kumar$^{2}$, Han Zhang$^{1}$, Emilio Martínez-Pañeda$^{3}$, Wei Tan$^{1*}$

\medskip

\begin{small}
$^{1}$ School of Engineering and Material Science, Queen Mary University of London, London, UK

$^{2}$ Material Science and Engineering, Delft University of Technology, Delft, Netherlands

$^{3}$ Department of Civil and Environmental Engineering, Imperial College London, London, UK\\
* Corresponding author (wei.tan@qmul.ac.uk)
\end{small}

\bigskip

\textbf{Abstract}

\vspace*{-2mm}

\end{center}

The project aims to explore a novel way to design and produce cellular materials with good energy absorption and recoverability properties. Spinodoid structures offer an alternative to engineering structures such as honeycombs and foam with scalability ensuring microscale benefits are reaped on a larger scale. Various materials and topologies have been utilised for numerical modeling and prototyping through additive manufacturing. Each design was evaluated using finite element modelling. Initial results from numerical models show anisotropic structures achieving high energy absorption efficiency. Through data-driven optimisation, results show a peak energy absorption value of 5.34 $MJ/m^{3}$ for anisotropic columnar structure. A physics-informed biased grid-search optimisation is faster due to parameters being explored in parallel. To validate the numerical model, compressive tests on various prototypes were conducted.

\begin{small}
\smallskip

\end{small}


\section{Introduction}
There has been significant progress in the advancement of architected materials over the past few decades with the central goal of developing lightweight, high performance materials. The use of energy-dissipating structures play an integral role in engineering and the rapid advancement in engineering fields such as transportation, aeronautics and automotive. They have led to an increased importance in the research of energy absorption capacity of structures. Examples of energy-dissipating structures used widely in engineering include crash boxes utilised in front bumper system in the automotive industry. Over the years, many studies focusing on various high specific energy absorbing materials and structures such as thin vessel structures and graded honeycombs have been conducted. Energy absorbers work by reducing the impact load distribution over a period of time and as such, an ideal material for use in impact protection systems would exhibit both increased energy absorption capability  and increased strength.  Through the use of optimal designs, lattice topologies can be constructed to have characteristics such as high strength, high damping, high stiffness, high deformability and low density \cite{guell2019}. The energy absorption of a material under deformation is defined as the integral of the resultant stress-strain response and can be optimised through high stress plateaus over a long strain regime \cite{jawaid2019}. Traditional thin walled prismatic structures are the industry standard, however, they offer limited energy absorption density during high impact scenarios. 

Recoverable cellular materials offer increased structural integrity due to their exceptional energy absorption and impact resistance. The main properties displayed by cellular structures involve being lightweight, reducing critical load, absorbing energy by irreversible means and having an increased specific energy absorption capacity \cite{galehdari2015}. Honeycombs, foams and fiber composite tubes are heavily utilised in various engineering applications as compared to monolithic solids they offer a compromise with increased stress plateaus enabling progressive failure through localized fracture, friction and instability. Spinodal structures offer an alternative to the high stress concentrations experienced traditional shell based structures at junctions as a result of their smooth topologies with minimised curvature \cite{alketan2019}. They are a group of shell based architectures which upon loading provide near consistent strain distribution, while also displaying fully homogenised behaviour and decreased imperfection sensitivity due to their stochastic nature \cite{guell2019}. The aim of this project is to provide a novel way to engineer the topology design and material selection so that an optimal design can be produced in terms of both energy absorption and recoverability. 

\section{Numerical Modelling}
Numerical models of isotropic and a number of anisotropic cases of spinodoid topologies were generated and investigated. The crystallographic texture of the spinodoid structures were controlled using a number of parameters represented by $\Theta$

\begin{equation}
    \Theta = (\rho,\theta_{1},\theta_{2},\theta_{3})
\end{equation}

\begin{figure}[h]
   \centering
   \includegraphics[width=1\textwidth]{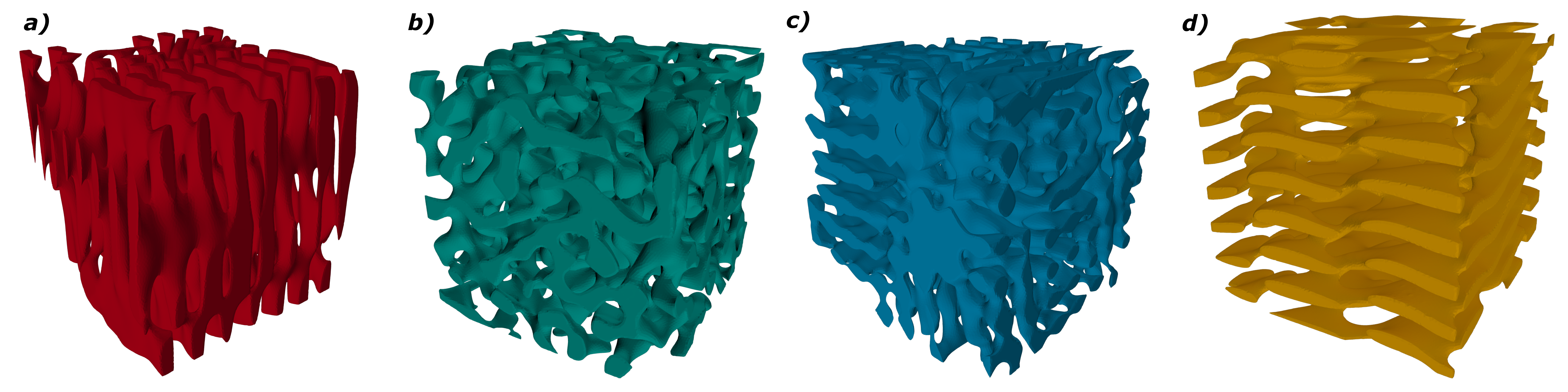}
   \caption{\label{fig:comb geom}Spinodoid topologies a) Columnar, b) Isotropic, c) Cubic and d) Lamellar.}
\end{figure}

With relative density ($\rho$) being defined as $\rho^{*}/\rho_{s}$, the ratio between the density of generated topology and homogenised solid, giving an indication of the structure porosity. The relative density of structures has been restricted to a range of [0.3, 1] as densities $\rho << 1$ result in discontinuous geometries, similar with angles being restricted to a minimum of 15\textdegree. Additionally, $\theta_{i}$ $\in$ $\{0\}$ $\cup$ [$\pi/12$,$\pi$/2] allows a fine control of grain orientation, with $\theta_{1} = \theta_{2} = 0$ and $\theta_{3} = \pi/2$ corresponding to an isotropic structure, as angles are restricted to a cone. In addition, values of $\theta_{3} < \pi/2$ results in the formation of Lamellar spinodoid structure. However, at $\theta_{3} = 0$ $\cap$ $\theta_{1,2} > 0$, results in the formation of Columnar structures. Finally, specific values of $\theta_{i}$ results in the formation of Cubic topology, each yielding a unique set of mechanical properties, shown in Figure \ref{fig:comb geom}. The materials utilised for finite element model are treated as isotropic homogenised continuum. Material properties of three different polymer types  were employed; Thermoplastic Polyurethane (TPU) being soft and ductile, Polyethylene Terephthalate Glycol (PET-G) a relatively brittle thermoplastic in comparison to TPU with high density and Carbon fibre reinforced PET-G (Carbon-P) being a composite; lightweight and stiff. To assess the failure of ductile materials, the Von Mises yield criterion is utilised. Capturing the elastic-plastic behaviour of spinodoid structures is essential in determining the energy absorption. Elastic region provides a recoverable method of absorbing energy, comparatively, plastic region has a larger potential for energy storage through plastic deformation, hence a trade-off between energy absorption and recoverability. The behaviour of numerical model undergoing high-strain compressive loading is assessed with a maximum strain of 50\%. To find the topological parameters that ensures highest energy absorption, data-driven optimisation approach is adopted, in tandem with preliminary testing of special topological cases.

\section{Results}
Preliminary numerical testing of design space revealed an affinity for anisotropic structures, that are parallel to the direction of loading. Cases such as the columnar hence performed significantly better than other topologies in terms of energy absorption with deformation primarily through buckling of columns, fracture and plastic deformation. Relative densities of 0.3, 0.5 and 0.7 were tested with energy absorption scaling with higher relative densities at the cost of heavier structure. It is not feasible to explore every combination of parameters provided by the large design space of inversely generated spinodoid structures. A guided, data-driven approach was adopted to find parameters that yielded exceptional energy absorption capability. This involved the use of prior knowledge, provided by preliminary testing. To reduce the design space needed to be explored, the consideration of relative density was omitted as high relative density resulted in greater energy absorption. Furthermore, as columnar spinodoid topologies resulted in higher energy absorption, hence $\theta_{3}$ was excluded. This allowed the design space to be optimised using two angles $\theta_{1,2}$. The optimisation of the two angles involved investigation of various samples of thetas from a range of [15,70]. From previous results, high $\theta_{1,2}$ results in isotropic material behaviour. 

\begin{figure}[h]
   \centering
   \includegraphics[width=1\textwidth]{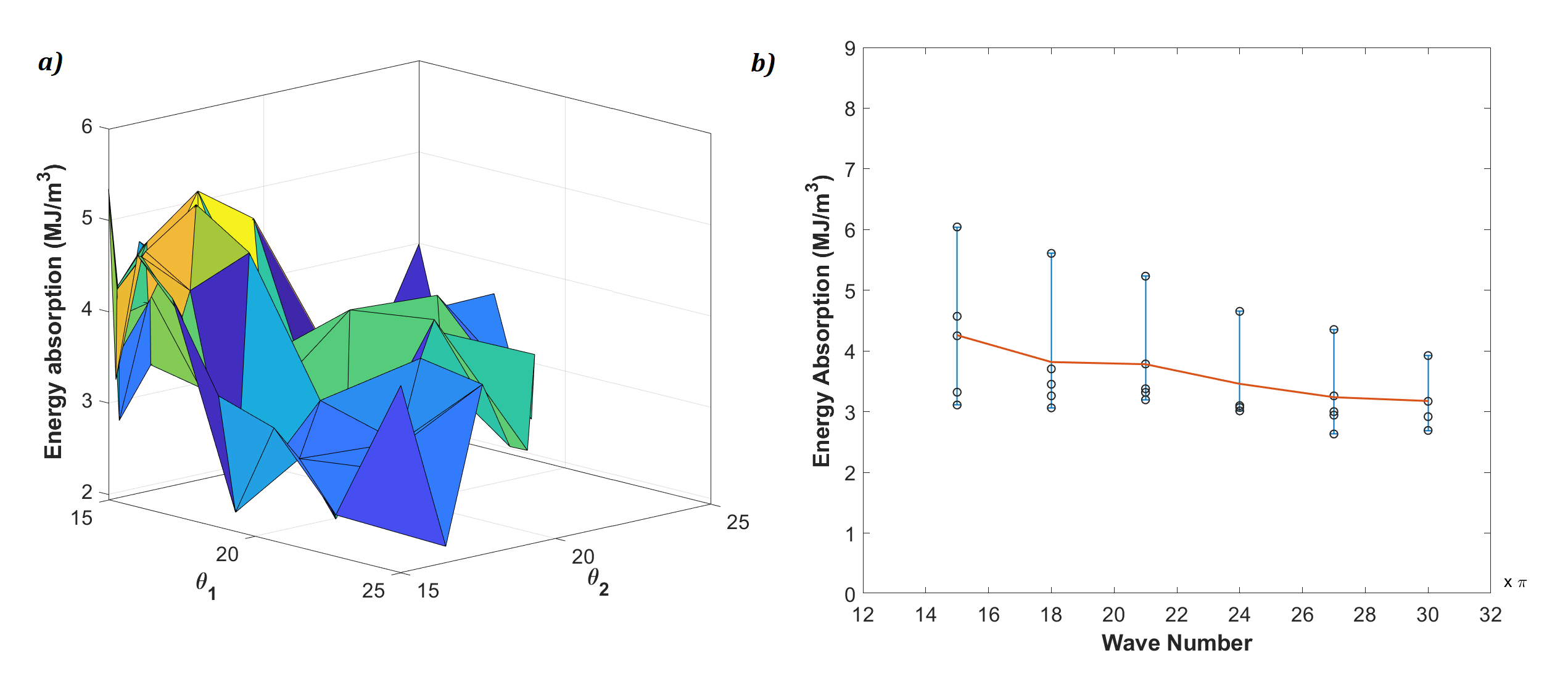}
    \caption{\label{fig:opt 15-25}a) Optimisation results for angles between 15 to 25 degrees on a polar grid and b) The convergence of energy absorption value over various wave numbers.}
\end{figure}

A biased sampling method on a polar coordinate grid was adopted to capture the behaviour of structure at low cone angles $\theta_{1,2}$, shown in Figure \ref{fig:opt 15-25}a). This revealed high energy absorption potential with angles in the range of [15,25]. Following the initial optimisation results, a further optimisation was completed within the theta range of [15.25], where angles closer to (15,15,0) resulted in higher energy absorption. The peak value of specific energy absorption was obtained to be approximately 5.34 $MJ/m^{3}$ at angles (15,15,0), with the effect of various angles on energy absorption also being shown by Figure \ref{fig:opt 15-25}a). To eliminate the stochastic nature of topology generation, higher wave numbers were utilised to obtain homogenised energy absorption value, with a converged value being obtained of 3.18 $MJ/m^{3}$ at wave number of 30$\pi$, shown in Figure \ref{fig:opt 15-25}b). The energy absorption efficiency of optimised spinodoid topologies was found to be of a larger magnitude than pre-existing structures - achieved at relatively low density. 



\section{Manufacturing and Experimental testing} 
Spinodal structures are typically manufactured through spinodal decomposition. This is when an epoxy template is formed by curing an epoxy and polyethylene glycol emulsion. It is then coated with the desired material and the epoxy template is removed through $O_2$ plasma ashing \cite{portela2020}. Whilst this manufacturing process is scalable, optimising the spinodoid for high energy absorption and low density is difficult due to the multitude of possible polymer ratios for the emulsion. Trying to reproduce a specific topology of spinodoid also proves challenging due to this natural process.

To avoid these issues, additive layer manufacturing methods were explored as the numerical model of the spinodoid could be reproduced indefinitely with consistent mechanical properties. Fused deposition modelling was used to prototype the numerical models whilst another additive layer manufacturing method such as selective laser sintering would then be used to scale up their production. Some features that were experimented with was the material and the scale of the spinodoid. There are three potential materials for the structure which are TPU, PET-G and Carbon Fibre-PET. Each material has a range of suitable temperatures to print with. Selecting an appropriate temperature for the spinodoid is important as it would reduce printing deformities such as stringing and extrusion issues. Printing the spinodoid topologies with each material allowed the material with the highest energy absorption to be identified.  Another consideration to have is the scale of the spinodoid. A structure with features smaller than the nozzle diameter will also produce printing issues such as under extrusion or failure to extrude at all. The minimum nozzle diameter available is 0.4mm, meaning the smallest feature on the spinodoid must be greater than 0.4mm to avoid any defects. Once the prototyping has concluded, an appropriate scalable manufacturing method can be used to mass-produce the structures. As the scale of the spinodoid has been identified for a nozzle diameter of 0.4mm, the structure can be scaled appropriately for different sized nozzles.

Initial tensile and compression tests for each material was done to verify the manufacturer’s given mechanical properties. Fused deposition modelling was used to print a solid dog bone test specimen for the tensile tests and a solid cube for the compression tests. The resulting data was used to improve the accuracy of the simulations by adjusting the material properties for their respective numerical model. Once each spinodal topology was printed out of the three materials, they were compressed until failure. This allowed the energy absorption for each structure to be measured and allowed the numerical simulations to be experimentally validated.

\section{Conclusions}

Spinodoid topologies exploits the double curvature orientation of surfaces, enhancing the entrapment of energy through plastic deformation resulting from a stiffer structure. When designed in optimal configuration, the architecture gives a specific energy absorption value of 5.34 $MJ/m^{3}$, higher than other pre-existing structures, such as the honeycombs or lattice-based structures. This is an optimistic finding in the potential of spinodoid topologies in the field of impact energy absorption. The method of data-driven modelling allows a supervised approach to find optimised topological parameters. Comparatively, this approach is faster than traditional sequential optimisation approach, where individual parameters are tested one at a time. Furthermore, multi-variable optimisation can also be considered as a way to fine-tune the optimised topology, depending on the application.

\subsection*{Acknowledgments}
I would like to thank Queen Mary's Apocrita HPC facility, supported by QMUL Research-IT and HPC facility at TU Delft for allowing us to run numerical simulations. W. Tan acknowledges financial support from the European Commission Graphene Flagship Core Project 3 (GrapheneCore3) under grant No. 881603.


\endgroup
\end{document}